\documentclass[twocolumn]{aastex701}

\usepackage{booktabs}
\usepackage{graphicx}

\begin{document}

\title{Line-driven Radiative Winds in B-Supergiants: Bridging the Gap between Fast and Slow m-CAK Solutions}

\author[orcid=0000-0001-9789-5098]{Matías Montesinos}
\affiliation{Departamento de Física, Universidad Técnica Federico Santa María, Avenida España 1680, Valparaíso, Chile}
\email[show]{matias.montesinosa@usm.cl}

\author[orcid=0009-0002-7943-8357]{Emil Zabala-Arroyo}
\affiliation{Departamento de Física, Universidad Técnica Federico Santa María, Avenida España 1680, Valparaíso, Chile}
\email[]{}

\author[orcid=0009-0009-5981-3640]{Juan José Castro-Salgado}
\affiliation{Departamento de Física, Universidad Técnica Federico Santa María, Avenida España 1680, Valparaíso, Chile}
\email[]{}

\author[orcid=0000-0002-2191-8692]{Michel Cur\'e}
\affiliation{Instituto de Física y Astronomía, Universidad de Valparaíso, Av. Gran Breta\~na 1111, Casilla 5030, Valpara\'iso, Chile}
\email[]{}

\author[orcid=0000-0002-8717-7858]{Ignacio Araya}
\affiliation{Centro Multidisciplinario de F\'isica, Vicerrector\'ia de Investigaci\'on, Universidad Mayor, 8580745 Santiago, Chile}
\email[]{}
\begin{abstract}
The modified Castor, Abbott, and Klein (m-CAK) theory predicts different wind regimes based on the line force parameter for changes in ionization ($\delta$) and the rotation parameter ($\Omega$). Stationary hydrodynamic studies have reported ``forbidden regions'' or gaps in this parameter space where no steady-state solution exists, suggesting physical instabilities. We investigate the stability of wind solutions within these gaps for B-supergiants to determine if they correspond to physical instabilities or numerical artifacts. We perform 1D time-dependent hydrodynamic simulations, systematically exploring the full $(\Omega, \delta)$ space for three B-supergiant models ($T_{\rm eff}=15-25$ kK), adopting a fixed density boundary condition. Our simulations reveal stable stationary solutions continuously across the entire parameter space, effectively filling the reported gaps. The transition from fast to slow regimes is smooth but structurally complex. Within the gap, the velocity profile develops a distinct ``kink'' or extended plateau in the supersonic flow, allowing the wind to reach a stable state. The mass-loss rate ($\dot{M}$) varies smoothly without artificial jumps. We find that the $\dot{M}$ gradient depends on the radiative driving strength ($k$): while $\dot{M}$ increases with $\delta$ for standard driving ($k \approx 0.32$), it decreases for the weak-driving regime ($k = 0.1$), consistent with stationary predictions. Moreover, in this regime, the final solution depends on the initial flow acceleration, confirming multiple hydrodynamic solutions. We conclude the m-CAK solution space is continuous; reported forbidden regions are artifacts of stationary methods. Time-dependent simulations effectively bridge the regimes, suggesting these transitions correspond to metastable states.

\end{abstract}

\keywords{
Hydrodynamics ---
Stellar winds ---
Stellar outflows ---
Massive stars ---
Stellar rotation ---
Numerical methods
}


\section{Introduction}
\label{sec:intro}

Radiation-driven winds from massive stars constitute one of the most critical mechanisms for mass and angular momentum loss throughout stellar evolution. The theoretical description of these outflows relies fundamentally on the modified Castor, Abbott, and Klein (m-CAK) theory \citep{cak-1975, m-cak1986, Pauldrach+1986}, a framework describing how momentum transfer from the stellar radiation field to the plasma via ultraviolet absorption lines accelerates material to supersonic velocities.

For massive stars, where rotation can play a significant role, the interplay between rotation and radiative acceleration introduces considerable complexity. Rotation modifies the effective gravity through centrifugal acceleration, reduces the wind velocity in equatorial regions, and can substantially enhance equatorial mass-loss rates due to the ``$\Omega$-slow'' solution mechanism \citep{Cure2004}. Early theoretical predictions demonstrated that rotating B[e]-supergiant winds exhibit marked latitude-dependent behavior, with dramatic density contrasts between polar and equatorial regions \citep{Pelupessy+2000}.

A critical theoretical challenge has emerged in recent years regarding the stability of these winds: the discovery of ``forbidden regions'' or gaps in the parameter space. Using the stationary hydrodynamic code \texttt{Hydwind} \citep{Cure2004}, \citet{Venero2016} demonstrated that for changes in the ionization parameter $\delta$, which accounts for the sensitivity of the line force to local density variations, and the rotation parameter $\Omega \equiv v_{\rm rot}(R_\star)/v_{\rm crit}$, no steady-state solution could be found within the standard m-CAK framework. These gaps were interpreted as regions of potential physical instability or abrupt bi-stability transitions \citep{Lamers+1995, Vink+1999}. However, recent multi-wavelength quantitative spectroscopy from the ULLYSES and XShootU collaborations challenges this view, finding no empirical evidence for an abrupt mass-loss jump in the temperature range of B-supergiants (e.g., \citealt{Verhamme+2024, deBurgos2024, Bernini-Peron2024}).

The existence of these forbidden regions raises fundamental questions: Are they manifestations of genuine physical instabilities in radiative wind dynamics, or do they constitute numerical artifacts arising from the limitations of stationary methods in resolving complex flow structure? Addressing these questions is crucial for predicting accurate mass-loss rates for stellar evolution and interpreting observations of rotating massive stars.

Subsequent work has provided important insights into the transition between fast and slow solutions. \citet{Araya+2018} performed numerical simulations to investigate the switching behavior between fast and $\Omega$-slow wind solutions in rapidly rotating Be-type stars, revealing that perturbations in the wind's base density can trigger transitions between different branches. More recently, analytical solutions for the $\delta$-slow regime were developed using closed-form velocity profiles \citep{Araya+2021}. Despite these advances, a systematic investigation of the physical nature of the forbidden region spanning the entire parameter space has remained elusive. Furthermore, observational studies of B-supergiants confirm the ubiquity of complex wind structures and intrinsic variability \citep{Markova+2008, Markova+2008b}. High-precision space photometry and long-term spectroscopy have revealed that photospheric pulsations can propagate into the wind, triggering dynamic, non-spherical outflows \citep{Aerts+2018, Simon-Diaz+2018A}, phenomena that stationary models inherently fail to capture.

Furthermore, recent stationary hydrodynamic studies by \citet{Venero+2024} have explored the ``weak driving'' regime characterized by low force multiplier values ($k \approx 0.1$). In this regime, the mass-loss rate exhibits an inverse dependence on ionization changes, decreasing $\dot{M}$ as $\delta$ increases, contrary to the standard high-$k$ behavior. Despite the significantly lower wind densities, stationary models still predict a forbidden region or gap in the parameter space. To provide a comprehensive stability analysis, we extend our study to include this weak driving scenario for the mid-B supergiant model, testing whether time-dependent simulations can resolve stable solutions even in these tenuous, low-momentum outflow regimes.

The key innovation of this work is the application of time-dependent hydrodynamic simulations to systematically explore the $(\Omega, \delta)$ parameter space without relying on stationary methods. Time-dependent simulations allow the wind to naturally evolve from initial conditions to a steady state, potentially revealing stable equilibrium solutions that stationary solvers cannot access due to algorithmic limitations (e.g., topology with multiple critical points).

In this study, we employ the \textsc{Fargo3d} hydrodynamical code \citep{Benitez-Llambay+2016} to perform 1D time-dependent simulations of rotating B-supergiant winds. Our specific aims are: 
(1) to determine the global structure of the solution space by systematically mapping the $(\Omega, \delta)$ plane, assessing whether stable stationary solutions exist within the previously identified forbidden regions; 
(2) to investigate the physical mechanism enabling stability by analyzing the velocity topology (e.g., kinks or plateaus) within the transition region; 
(3) to verify the robustness of these solutions over extended dynamical timescales; and 
(4) to clarify the role of photospheric boundary conditions in the mass-loss rate determination, specifically investigating how these conditions influence the emergence of the abrupt jumps in mass loss reported in stationary models.

The paper is organized as follows: \S~\ref{sec:method} describes our numerical methodology and physical model. \S~\ref{sec:results} presents the global parameter space maps, the discovery of solutions within the gap, and the stability analysis of the weak driving regime. \S~\ref{sec:discussion} discusses the physical implications, the flow structure, and the discrepancy with stationary mass-loss predictions. Finally, \S~\ref{sec:conclusions} summarizes our conclusions.

\section{Numerical method and physical model}
\label{sec:method}

To model the time-dependent evolution of the radiation-driven stellar wind, we employ the hydrodynamical code \texttt{Fargo3D} \citep{Benitez-Llambay+2016}, in which we implemented the m-CAK line-acceleration terms and specific boundary conditions required for an equatorial stellar wind outflow. The setup consists of a one-dimensional (1D) radial domain, assuming full azimuthal and latitudinal symmetry.

\subsection{Governing equations}

We solve the time-dependent hydrodynamic equations for mass and momentum conservation in the radial direction. The gas is treated as an inviscid, isothermal fluid. The governing equations are:

The mass conservation:
\begin{equation}
    \frac{\partial \rho}{\partial t} + \frac{1}{r} \frac{\partial}{\partial r} (r \rho v_r) = 0,
\end{equation}

And the radial momentum conservation:
\begin{equation}
    \frac{\partial v_r}{\partial t} + v_r \frac{\partial v_r}{\partial r} - \frac{v_\phi^2}{r} = -\frac{1}{\rho}\frac{\partial P}{\partial r} - \frac{GM_\star(1 - \Gamma_e)}{r^2} + g_{\mathrm{line}}.
\end{equation}

Here, $\rho$ is the mass density, $v_r$ is the radial velocity, $v_\phi$ represents the azimuthal velocity component, and $P$ is the gas pressure. The term $\Gamma_e$ is the Eddington factor for electron scattering, defined as $\Gamma_e \equiv \sigma_e L_\star/4 \pi c G M_\star$, where $\sigma_e$ is the electron scattering opacity, $L_\star$ is the stellar luminosity, $M_\star$ is the stellar mass, $G$ is the gravitational constant, and $c$ is the speed of light. This term accounts for the radiation pressure from continuum scattering, which effectively reduces the stellar gravity to an effective value $g_{\mathrm{eff}} = g(1 - \Gamma_e)$, where $g = GM_*/r^2$. 

To close the system of equations, we adopt an isothermal equation of state, $P = c_s^2 \rho$, where $c_s$ is the constant isothermal sound speed, defined as the proton thermal speed $c_s = \sqrt{2 k_B T_{\rm eff} / m_p}$, with $k_B$ being the Boltzmann constant, $m_p$ the proton mass and $T_{\rm eff}$ is the stellar effective temperature. This choice implies that the gas pressure depends solely on the density distribution, neglecting any radial gradients in the gas temperature \citep[see for instance][]{Gormaz-Matamala2021}. Furthermore, it is important to note that the gas pressure force formally depends on the gas kinetic temperature ($T_{\rm gas}$), which is not necessarily equal to $T_{\rm eff}$ and cannot be self-consistently computed within our current framework. This approximation, where $T_{\rm gas} \approx T_{\rm eff}$ is assumed for the entire wind.

\subsection{The radiation force: m-CAK implementation}

The wind is driven by the line acceleration $\mathbf{g}_{\mathrm{line}}$, calculated following the m-CAK theory. In our implementation, this term is given by:

\begin{equation}
    g_{\mathrm{line}} = \frac{\sigma_e L_\star}{4 \pi c r^2} \, M(t) \, f_{\mathrm{disk}}(r, v_r, v_r'),
\end{equation}
where $\sigma_e$ is the electron scattering opacity, $L_\star$ is the stellar luminosity, and $c$ is the speed of light. The force multiplier $M(t)$ is defined as:

\begin{equation}\label{eq:force_multiplier}
    M(t) = k \, t^{-\alpha} \left( \frac{n_e}{W(r)} \right)^\delta,
\end{equation}
where $k, \alpha,$ and $\delta$ are the force multiplier parameters. Here, $n_e$ is the electron number density and $W(r)$ is the geometric dilution factor, given by:

\begin{equation}
    W(r) = \frac{1}{2} \left( 1 - \sqrt{1 - \left(\frac{R_\star}{r}\right)^2} \right).
\end{equation}

The dimensionless optical depth parameter $t$ is computed using the local density and velocity gradient \citep{cak-1975}:

\begin{equation}
    t = \frac{\sigma_e \rho v_{\mathrm{th}}}{|dv_r/dr|},
\end{equation}
where $v_{\mathrm{th}}$ is the proton thermal speed.

The finite disk correction factor, $f_{\mathrm{disk}}$, accounts for the non-point source nature of the stellar radiation field \citep{m-cak1986, Pauldrach+1986}. Following the standard formalism, it is given by:

\begin{equation}
    f_{\mathrm{disk}} = \frac{(1+\sigma)^{1+\alpha} - (1+\sigma\mu_\star^2)^{1+\alpha}}{(1+\alpha)\sigma(1+\sigma)^\alpha (1-\mu_\star^2)},
\end{equation}
where $\mu_\star = \sqrt{1 - (R_\star/r)^2}$ is the cosine of the finite cone angle subtended by the star, and the function $\sigma$ is defined as:

\begin{equation}
    \sigma = \frac{d \ln v_r}{d \ln r} - 1 = \frac{r}{v_r}\frac{dv_r}{dr} - 1.
\end{equation}

A detailed presentation of the stationary m-CAK formalism, including the structure of the equation of motion, solution topology, and boundary conditions, is given in \cite{Cure_review_2023}.

\begin{table*}[t]
\caption{Stellar and wind parameters for the representative models. Mass ($M_\star$), luminosity ($L$), and the Eddington parameter ($\Gamma_e$) are derived from the input $T_{\rm eff}$, $\log g$, and $R_\star$ following the standard relations for an ionized hydrogen atmosphere (see Sect. \ref{sec:method}).}
\label{tab:stellar_params}
\centering
\begin{tabular}{l c c c c c c c c}
\hline\hline
Model & $T_{\rm eff}$ (K) & $\log g$ & $R_\star$ ($R_\odot$) & $M_\star$ ($M_\odot$) & $\log (L/L_\odot)$ & $\Gamma_e$ & $\alpha$ & $k$ \\
\hline
T15   & 15,000 & 2.11 & 52 & 12.70 & 5.09 & 0.25 & 0.50 & 0.32 \\
T19   & 19,000 & 2.50 & 40 & 18.45 & 5.27 & 0.27 & 0.50 & 0.1 -- 0.32 \\
T25   & 25,000 & 2.90 & 35 & 35.49 & 5.63 & 0.32 & 0.55 & 0.34 \\
\hline
\end{tabular}
\end{table*}

\subsection{Numerical domain and initial conditions}\label{sec:methods}

To ensure high numerical precision and accurately resolve the complex flow structures within the transition region, the computational domain is discretized using a high-resolution 1D logarithmic grid consisting of $N_r = 2048$ radial cells. The grid extends from the stellar surface at $R_{\mathrm{in}} = R_\star$ to an outer boundary at $R_{\mathrm{out}} = 100 R_\star$. 

The system is initialized using a $\beta$-velocity law profile, $v_r(r) = v_{\infty}(1 - R_\star/r)^\beta$. For the standard stellar models ($k \approx 0.32$), we adopted $\beta = 1.0$ for all simulations. However, for the weak driving regime ($k = 0.1$), we observed that the final steady state is slightly influenced by the initial acceleration profile (see Appendix \ref{sec:appendix_snapshots}). To consistently reproduce the distinct ``fast'' and ``slow'' solution branches reported by \citet{Venero+2024}, we adopted branch-specific initial conditions: $\beta = 1.0$ was used to access the fast regime, while $\beta = 2.0$ was required to access the transition and slow wind regions.

The initial density profile is derived from mass conservation, $\rho(r) = \dot{M} / (4\pi r^2 v_r)$. The azimuthal velocity is initialized as $v_\phi(r) = v_{\mathrm{rot}}(R_\star/r)$, where $v_{\mathrm{rot}}$ represents the rotation speed at the stellar surface ($r=R_\star$) at the equator. We define the rotation rate via the dimensionless parameter $\Omega = v_{\mathrm{rot}} / v_{\mathrm{crit}}$, where $v_{\mathrm{crit}}$ is the critical breakup velocity. This is calculated as $v_{\mathrm{crit}} = \sqrt{GM_\star(1-\Gamma_e)/R_\star}$, which corresponds to $v_{\mathrm{crit}} = v_{\mathrm{esc}}/\sqrt{2}$, where $v_{\mathrm{esc}}$ is the effective escape velocity (e.g., \citealt{Puls+2008}.)

Regarding the boundary conditions, we explicitly set the photosphere as the inner boundary ($r=R_*$), following the approach of \citet{Venero2016}. At this boundary, we implement a fixed-value condition by prescribing a constant mass density $\rho_0 = 10^{-11} \text{ g cm}^{-3}$ and a subsonic injection velocity $v_0 = c_s/4$ for all models in our grid. It is important to note that $\rho_0$ is an independent input parameter prescribed to ensure a stable subsonic base, rather than a value derived from the injection velocity. To physically characterize this boundary placement, we calculated the corresponding Rosseland optical depth ($\tau_{\rm Ros}$), finding that $\tau_{\rm Ros}(R_*)$ ranges from approximately 0.086 to 0.132 across our model grid (with $\tau_{\rm Ros} \approx 0.096$ for the T19 case). These values indicate that our numerical domain begins in the optically thin regime. Following \citet{Araya+2018}, this fixed-density approach in the thin regime is a necessary configuration to ensure a stable subsonic reservoir, allowing the wind to relax self-consistently towards a steady state while avoiding the numerical instabilities at the base that often occur when the connection point is placed in much deeper, optically thick layers (e.g., \citealt{Josiek2025}). This configuration allows the wind to accelerate and pass through the sonic point further out in the domain without artificial feedback from the sub-photospheric structure.

\subsection{Stellar models}
\label{sec:stellar_models}

To investigate the stability of m-CAK solutions and facilitate a direct comparison with previous stationary hydrodynamic studies, we adopt the stellar parameters defined by \citet{Venero2016}. We selected three representative models that span the B-supergiant spectral range: T15 (late-type, $T_{\mathrm{eff}}=15\,000$ K), T19 (mid-type, $T_{\mathrm{eff}}=19\,000$ K), and T25 (early-type, $T_{\mathrm{eff}}=25\,000$ K).

The physical properties and the corresponding line-force parameters ($\alpha, k$) for each model are listed in Table~\ref{tab:stellar_params}. We explore the parameter space by performing a total of 567 time-dependent hydrodynamic simulations. For each stellar configuration, we vary the rotation rate over a grid of nine values, $\Omega \in \{0, 0.1, 0.2, 0.3, 0.4, 0.5, 0.6, 0.7, 0.8, 0.9\}$, and the ionization parameter $\delta$ in the range $[0.0, 0.40]$ with a uniform resolution of $\Delta \delta = 0.02$.

Regarding the chemical composition, we adopt solar abundances for all models, with a helium-to-hydrogen number ratio of $N_{\rm He}/N_{\rm H} = 0.1$. For the temperature range of our B-supergiant grid ($15,000$ to $25,000$ K), we assume the wind remains fully ionized, resulting in a mean molecular weight of $\mu \approx 0.62$ and a fixed electron scattering opacity of $\kappa_e = 0.34$ cm$^2$ g$^{-1}$. This opacity is used to define the electron Eddington factor, $\Gamma_e = \kappa_e L / (4 \pi G M c)$, which characterizes the radiative support against gravity at the base of the wind.

\begin{figure*}
    \centering
    \includegraphics[width=0.84\textwidth]{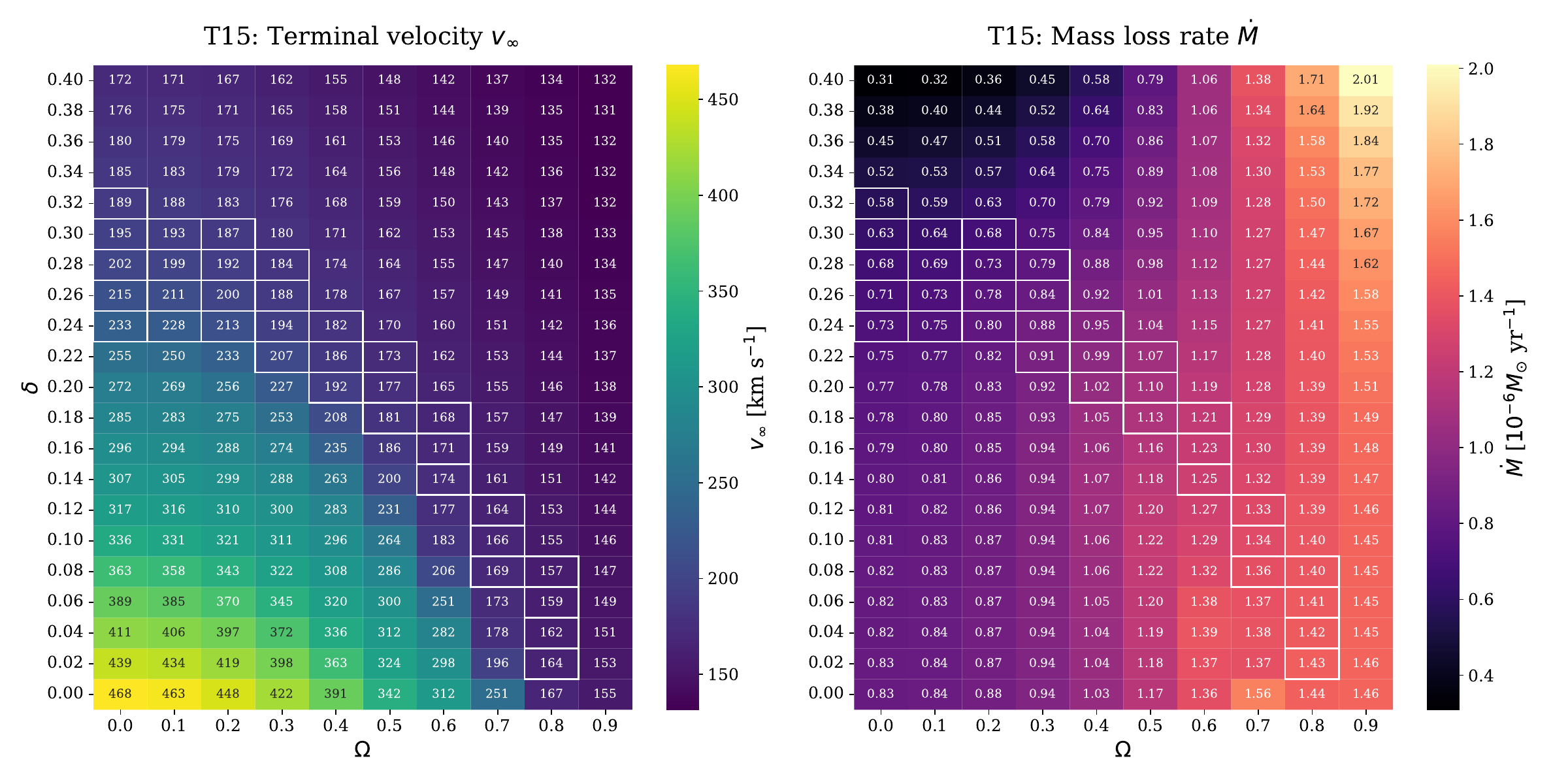}
    
    \includegraphics[width=0.84\textwidth]{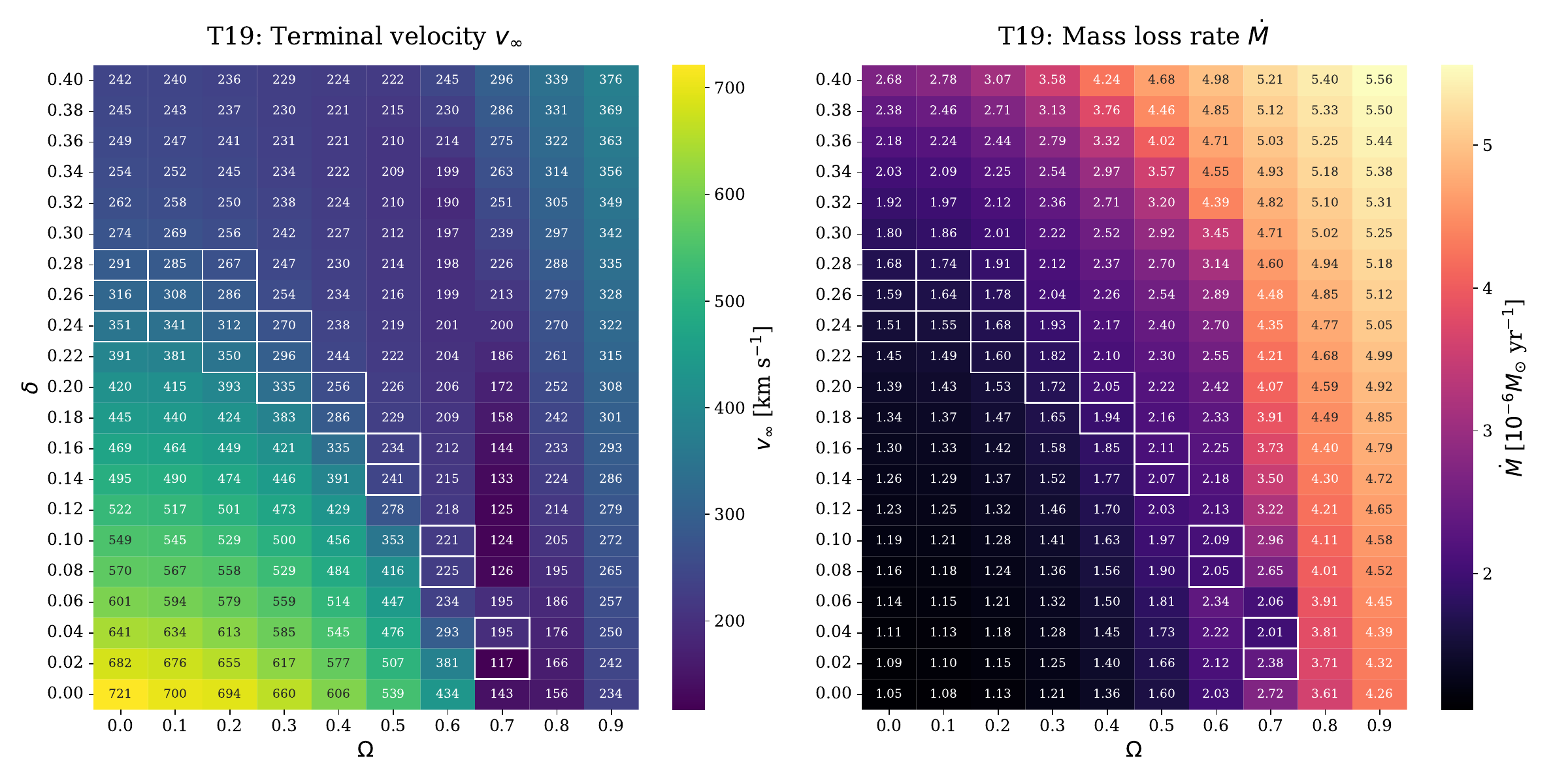}
    
    \includegraphics[width=0.84\textwidth]{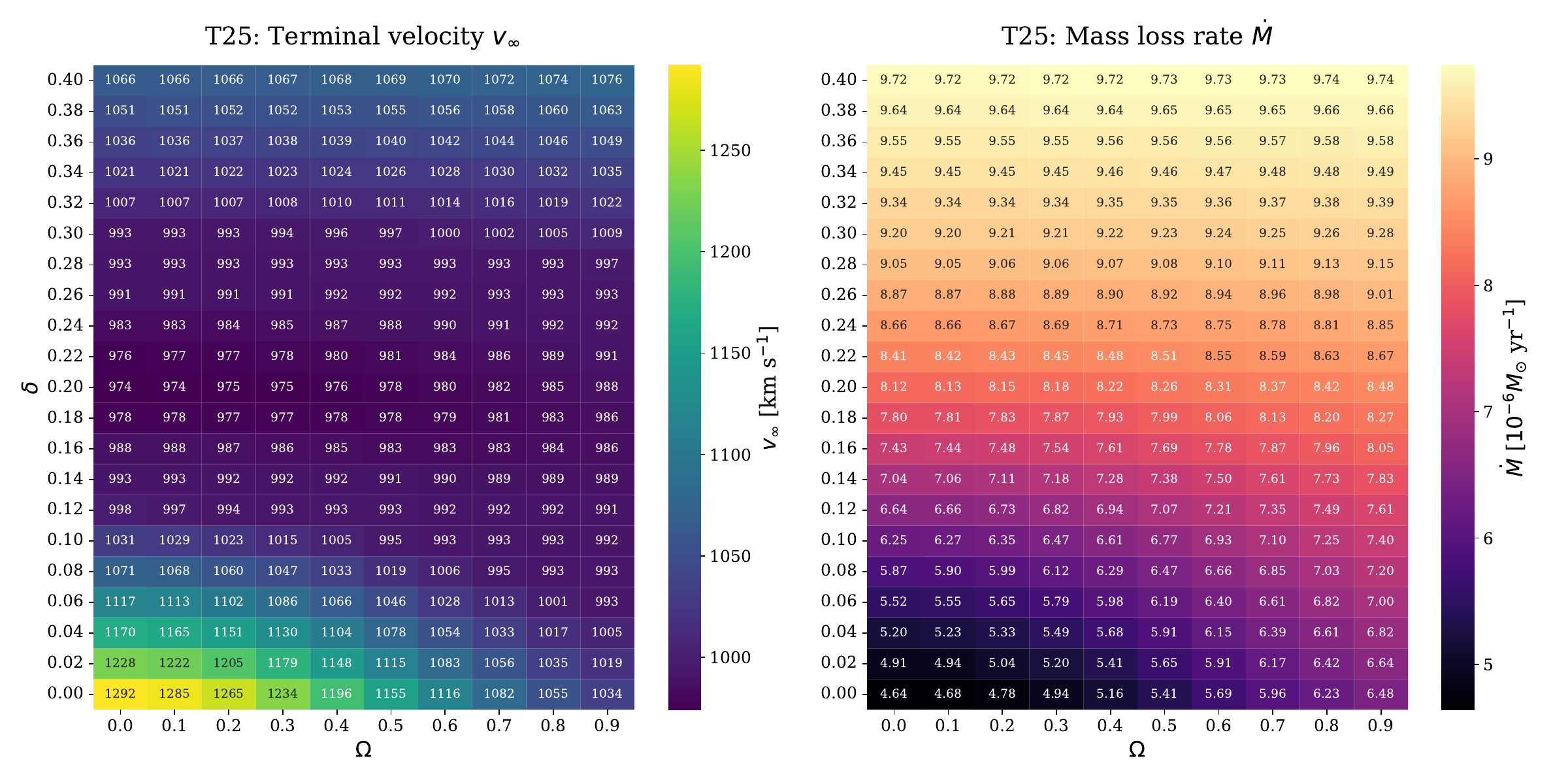}
    
    \caption{Maps of terminal velocity ($v_\infty$, left panels) and mass-loss rate ($\dot{M}$, right panels) in the $(\Omega, \delta)$ parameter space. From top to bottom: Model T15 (late B-supergiant), T19 (mid B-supergiant), and T25 (early B-supergiant). The color gradient indicates the magnitude of the stable stationary solution found by our time-dependent simulations. The white solid rectangles overlaid on the T15 and T19 maps indicate the ``forbidden regions'' (gaps) reported in previous stationary studies \citep{Venero2016}, where no solution was found. Our results demonstrate continuous stable solutions across the entire domain, filling these gaps.}
    \label{fig:heatmaps_global}
\end{figure*}

\section{Results}\label{sec:results}

We have performed a systematic exploration of the hydrodynamic solutions for radiation-driven winds across the $(\Omega, \delta)$ parameter space. For each of the three representative stellar models described in \S~\ref{sec:stellar_models} (T15, T19, and T25), we conducted time-dependent simulations covering the full range of rotation rates $\Omega \in [0.0, 0.9]$ and ionization parameters $\delta \in [0.00, 0.40]$, evolving each model until a steady state was achieved. To facilitate a direct comparison with the stationary results of \cite{Venero2016}, the force multiplier parameters ($\alpha$ and $k$) were kept fixed for each stellar model, adopting the specific values listed in Table~\ref{tab:stellar_params}.

\subsection{Global existence of stationary solutions}

The steady-state solutions obtained from our time-dependent simulations reveal a continuous distribution of wind parameters across the entire $(\Omega, \delta)$ domain. Figure~\ref{fig:heatmaps_global} displays the resulting terminal velocity ($v_\infty$) and mass-loss rate ($\dot{M}$) maps for the three stellar models. The white solid rectangles overlaid on the T15 and T19 maps indicate the ``forbidden regions'' (gaps) reported by \citet{Venero2016}, where the stationary \texttt{Hydwind code} \citep{Cure2004} failed to converge to a solution.

The most prominent feature of these maps is the continuous presence of stable solutions. Regardless of the spectral type or rotation rate, our time-dependent approach finds a physical steady-state solution throughout the entire analyzed domain, including the regions previously identified as unstable gaps.

For the late and mid B-supergiants (Models T15 and T19), the terminal velocity decreases monotonically as $\delta$ increases, transitioning smoothly from the fast wind regime to the slow wind regime without any discontinuity. The color gradients in Fig. ~\ref{fig:heatmaps_global} show that this transition occurs precisely across the hatched regions. This suggests that the ``gap'' is not a region devoid of solutions, but rather a domain where the flow structure becomes complex---likely involving multiple critical points or multiple coexisting solutions preventing stationary models from converging.

Regarding the mass-loss rate, the simulations reveal a strong dependence on the stellar parameters, the rotation rate, and the change in ionization through the wind parameter $\delta$. For the standard T15 model ($k=0.32$), $\dot{M}$ varies smoothly within the $10^{-6} M_{\odot} \text{ yr}^{-1}$ regime. At $\Omega=0$, it decreases moderately from $0.83$ to $0.31 \times 10^{-6} M_{\odot} \text{ yr}^{-1}$ as $\delta$ increases, whereas for high rotators ($\Omega=0.9$) it rises to $2.01 \times 10^{-6} M_{\odot} \text{ yr}^{-1}$. This behavior is consistent across the standard models, where the radiative driving remains efficient enough to sustain a substantial outflow.

For the early B-supergiant (Model T25), previous studies \citep{Venero2016} reported only a fast solution branch. Our simulations (bottom panels of Fig. ~\ref{fig:heatmaps_global}) confirm the existence of stable solutions up to $\delta=0.40$. However, an important physical difference is observed: even at high $\delta$, the wind velocity remains relatively high ($v_\infty \gtrsim 1000$ km s$^{-1}$). This implies that for high-luminosity stars, the radiative force is strong enough to prevent the wind from dropping into a truly ``slow'' regime ($v_\infty < 500$ km s$^{-1}$) within the standard m-CAK parameter space.

\subsection{Validation and continuity of the transition}
\label{sec:validation}

To validate our time-dependent implementation and quantify the behavior of the wind across the transition from the fast to the slow regime, we compare our results directly with the stationary solutions reported by \cite{Venero2016} for the standard model T19 ($\Omega=0.0$).

\begin{figure}
    \centering
    \includegraphics[width=\hsize]{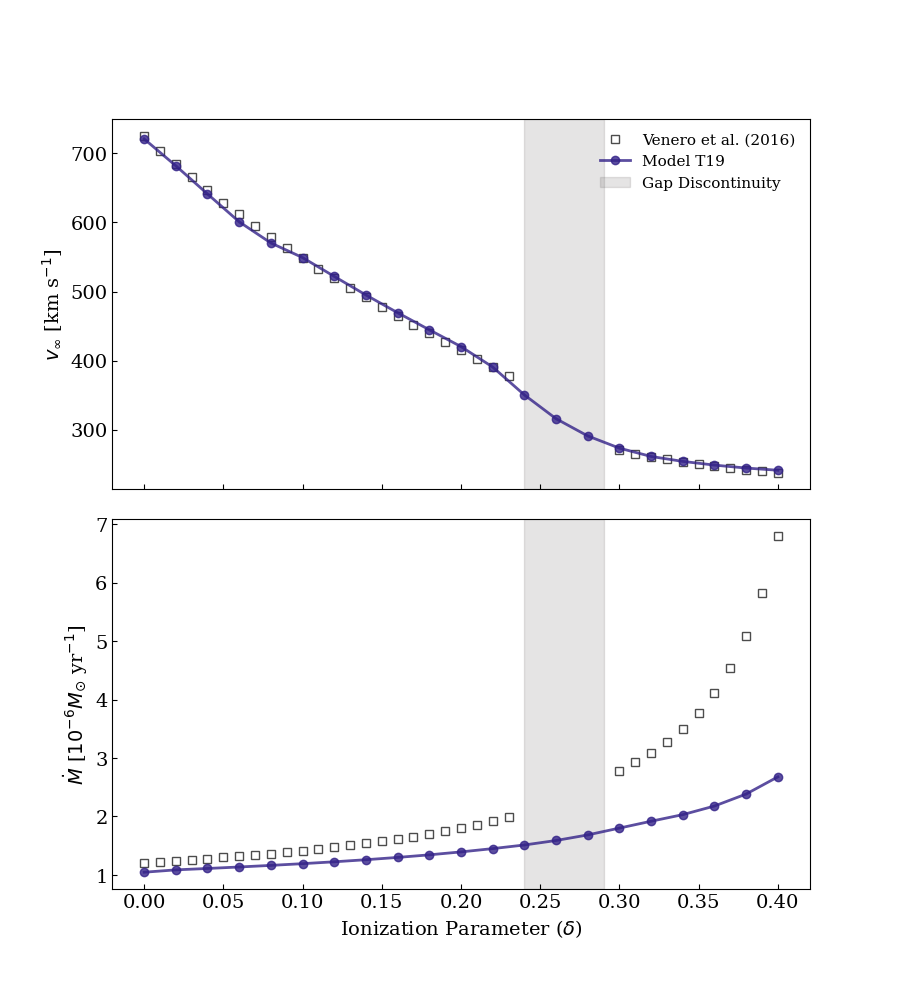}
    
    \caption{Quantitative comparison between our time-dependent simulations (blue lines/circles) and the stationary results from \cite{Venero2016} (black squares) for model T19 with no rotation ($\Omega=0.0$). \textit{Top panel:} Terminal velocity vs. ionization parameter $\delta$. \textit{Bottom panel:} Mass-loss rate vs. $\delta$. The gray hatched area indicates the ``gap'' reported in the stationary study. Note the inflection in the velocity curve across the gap and the bounded increase in $\dot{M}$ in the slow wind regime, contrasting with the stationary jump (see text for discussion).}
    \label{fig:comparison_t19}
\end{figure}

Figure~\ref{fig:comparison_t19} presents a comparison of the terminal velocity ($v_\infty$, top panel) and mass-loss rate ($\dot{M}$, bottom panel) as a function of the ionization parameter $\delta$. The stationary data from \cite{Venero2016} are shown as squares, while our time-dependent results are depicted by solid lines.

Across the entire parameter space, including both the fast ($\delta \le 0.20$) and slow ($\delta \ge 0.30$) wind regimes, we find an exact numerical correspondence with the stationary predictions. The terminal velocity values are identical to the m-CAK results, confirming that our time-dependent implementation accurately recovers the stationary physics. This perfect convergence validates that the adopted photospheric injection velocity ($v_0 = c_s/4$) and fixed-density boundary effectively synchronize the time-dependent flow with the m-CAK regularity conditions.

In the region previously identified as a gap ($0.21 \le \delta \le 0.28$, hatched area), the terminal velocity follows a continuous path that matches the theoretical m-CAK branch with high precision. The inflection or ``change of rhythm'' observed in the $v_{\infty}$ curve (Fig.~\ref{fig:comparison_t19}, top panel) is therefore an intrinsic property of the m-CAK solution topology and not a numerical artifact. Our time-dependent approach shows that this transition is hydrodynamically stable and can be reached continuously, bridging the fast and slow regimes and resolving the convergence issues inherent to stationary methods. While stationary models fail to converge here, likely due to the presence of multiple critical points or solution discontinuities, the time-dependent approach successfully finds a stable ``bridge'' solution that connects the fast and slow regimes.

A fundamental physical difference appears in the mass-loss rate behavior within the slow wind regime ($\delta \gtrsim 0.30$). As shown in the bottom panel of Fig.~\ref{fig:comparison_t19}, the stationary models predict a sudden, exponential increase (``jump'') in $\dot{M}$ immediately after the gap. In contrast, our time-dependent results reveal a moderate, monotonic increase in $\dot{M}$. While the stationary predictions inflate $\dot{M}$ by a factor of $\sim 4$ across the transition, our simulations show a much gentler rise (from $\sim 1.10$ to $2.37 \times 10^{-6} M_{\odot} {\rm yr}^{-1}$). This discrepancy arises from the boundary conditions. The stationary \texttt{Hydwind} code typically fixes the optical depth ($\tau_\star$), forcing an artificial density increase at the base when velocity gradients are small. By fixing the photospheric density ($\rho_0$)---which following \citet{Araya+2018} is a more physically robust condition for a stable subsonic reservoir---the mass flux is determined self-consistently by the sonic point requirements. This results in a stable and bounded increase in mass loss rather than an artificial runaway.

\subsection{Topological mechanism and structural consistency}
\label{sec:topology}

\begin{figure}
    \centering
    \includegraphics[width=\hsize]{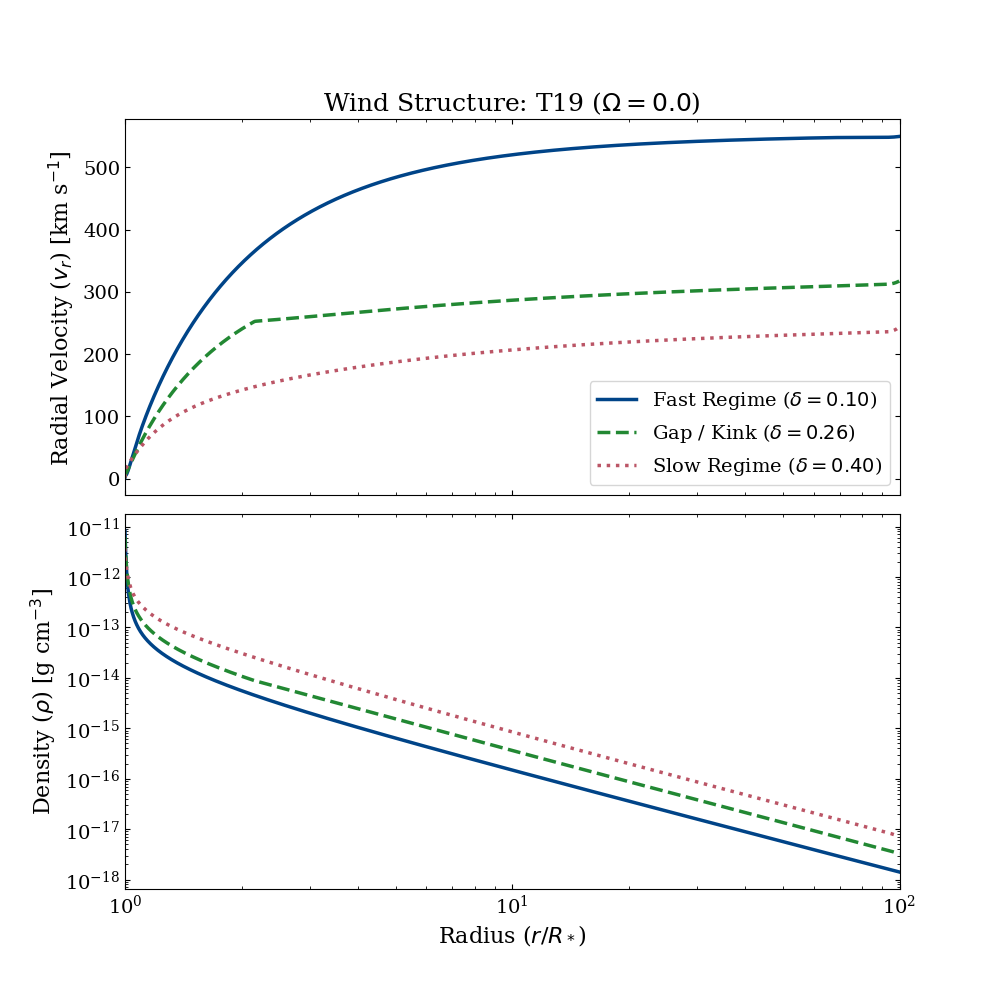}
    \caption{Wind structure for model T19 ($\Omega=0.0$) in the fast regime ($\delta=0.10$, blue solid line), gap region ($\delta=0.26$, green dashed line), and slow regime ($\delta=0.40$, red dotted line). Top panel: Radial velocity profiles. The solution within the gap exhibits a velocity plateau or ``kink'' extending from $2\,R_\star$ to $10\,R_\star$. Bottom panel: Mass density profiles (log scale). The density profiles are consistent with their corresponding velocity regimes, where lower velocities result in higher density levels as required by mass conservation.}
    \label{fig:topology_structure}
\end{figure}

To investigate the physical mechanism underlying the stability of the solutions within the so-called ``forbidden'' regions, we analyze the radial structure of the wind. Figure~\ref{fig:topology_structure} displays the velocity profiles $v_r(r)$ (top panel) and the density distribution $\rho(r)$ (bottom panel) for model T19 ($\Omega=0.0$) corresponding to three representative ionization states: the fast regime ($\delta=0.10$, blue solid line), the gap region ($\delta=0.26$, green dashed line), and the slow regime ($\delta=0.40$, red dotted line).

The fast solution ($\delta=0.10$) exhibits the standard convex topology typical of m-CAK winds. Its velocity profile undergoes a rapid acceleration near the stellar surface ($r \lesssim 2 R_\star$) and smoothly transitions into an asymptotic flattening around $r \approx 10 R_\star$ as it approaches the terminal velocity. Consequently, the density profile follows the expected steep decline driven by both spherical expansion and rapid acceleration ($\rho \propto r^{-2}v^{-1}$).

In contrast, the solution within the gap ($\delta=0.26$) reveals a complex, hybrid structure. As shown in the top panel of Fig. ~\ref{fig:topology_structure}, the wind initially accelerates near the photosphere but encounters a distinct velocity ``kink'' in the region $2\,R_\star \lesssim r \lesssim 10\,R_\star$. In this region, the radiative acceleration is locally suppressed by the $(n_e/W)^\delta$ term in the force multiplier (Eq.~\ref{eq:force_multiplier}). Since $n_e \propto \rho$, this term introduces a dependency $g_{\rm line} \propto \rho^\delta$, which reduces the driving efficiency as the wind density drops, causing the observed stagnation before the final acceleration phase.

This kinematic feature has a direct hydrodynamic counterpart in the density structure (bottom panel). In the region of the velocity plateau, where $\partial v_r / \partial r \approx 0$, the density fall-off becomes purely geometric ($\rho \propto r^{-2}$). The resulting density profiles remain consistent with their respective velocity regimes; to satisfy mass conservation, the lower expansion velocities in the gap and slow regimes lead to higher mass densities throughout the wind compared to the fast solution.

Finally, the slow solution ($\delta=0.40$, red dotted line) represents a distinct stable regime. Unlike the gap solution, it does not exhibit a velocity plateau. Instead, it shows a monotonic but significantly shallower acceleration profile throughout the entire domain. The terminal velocity is low ($v_\infty \approx 260$ km s$^{-1}$), and the density remains consistently higher than in the fast and gap regimes due to the lower velocity. This confirms that the ``slow'' branch is a robust, continuous solution where the radiative driving is less efficient but stable.

This inflection within the gap, characterized by a deceleration or stagnation followed by re-acceleration, likely mimics a singularity in stationary momentum equations, causing standard shooting methods to fail. While our time-dependent approach naturally evolves the flow into this configuration, the emergence of such a complex topology suggests that the wind may be residing in a metastable state associated with multiple stable branches. This finding hints at a richer hydrodynamic landscape than previously assumed, warranting further investigation to fully characterize the potential coexistence of multiple attractors in this transition region.

\subsection{The weak driving regime ($k=0.1$)}
\label{sec:weak_driving}

To complement our analysis, we extended our simulations to the ``weak driving'' regime ($k=0.1$) using the T19 model. Recent stationary studies \citep{Venero+2024} showed that for such low force multiplier values, the wind properties change drastically, with the mass-loss rate decreasing as changes inionization increases.

\begin{figure*}
   \centering
   \includegraphics[width=\textwidth]{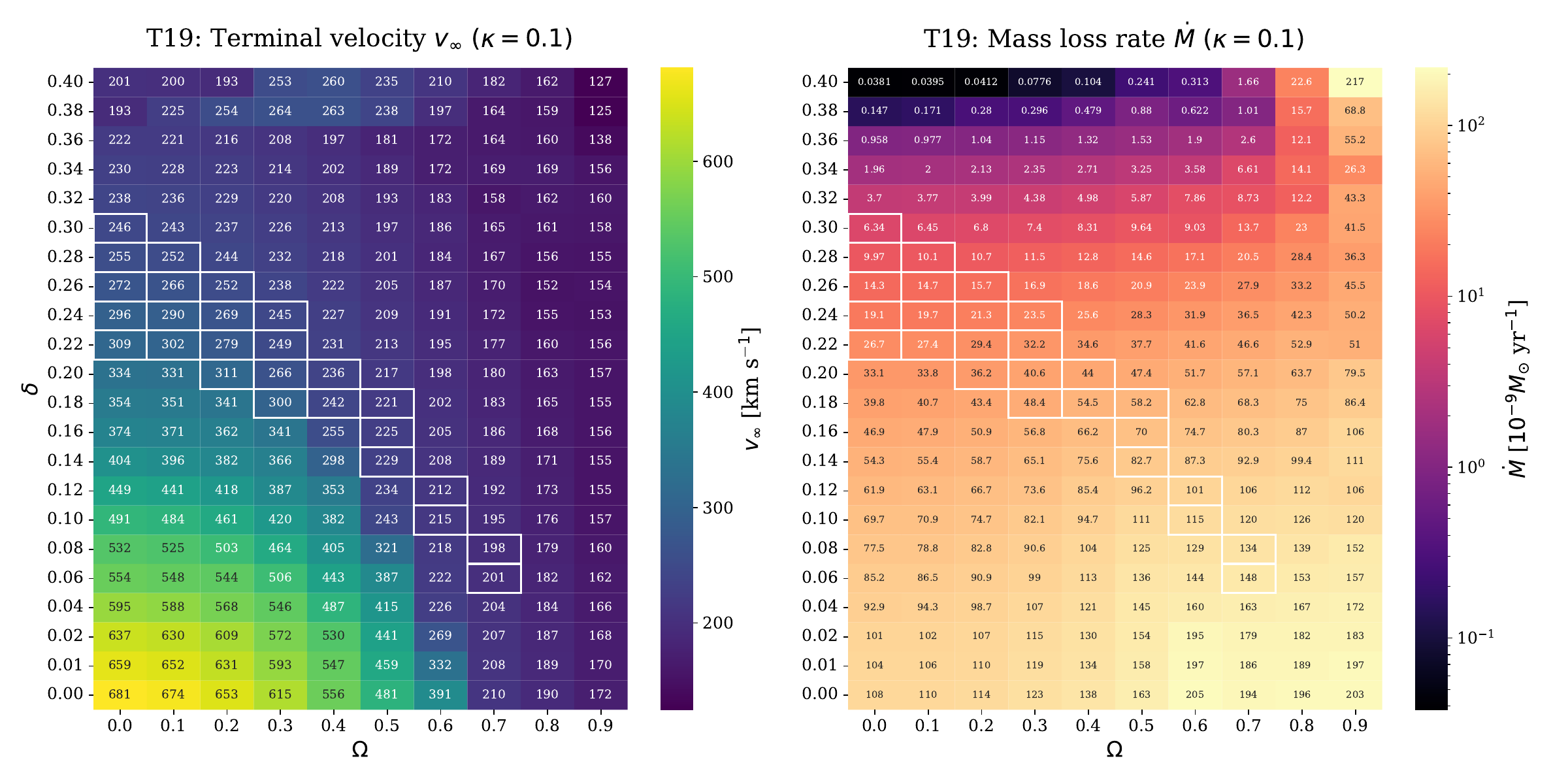}
   \caption{Global existence maps for the weak driving regime (Model T19 with $k=0.1$). \textit{Left panel:} Terminal velocity $v_\infty$. \textit{Right panel:} Mass-loss rate $\dot{M}$. Unlike the standard models, $\dot{M}$ decreases drastically as $\delta$ increases, dropping by orders of magnitude into a ``dead wind'' state for $\delta > 0.30$. However, stable solutions are found across the entire domain.}
   \label{fig:k01_results}
\end{figure*}

Figure~\ref{fig:k01_results} (left panel) presents the global stability map for this regime. Our time-dependent code finds stable stationary solutions across the entire parameter space, filling the ``forbidden regions'' reported in stationary works. The terminal velocity decreases monotonically with $\delta$, similar to the standard case.

However, the mass-loss rate behavior (Fig.~\ref{fig:k01_results}, right panel) confirms the theoretical inversion predicted for low-$k$ winds. As $\delta$ increases, $\dot{M}$ drops significantly. For the fast wind ($\delta \lesssim 0.2$), $\dot{M}$ is in the range of $10^{-7} M_{\odot} \text{ yr}^{-1}$, but it falls exponentially to orders of magnitude below $10^{-9}$ in the slow regime ($\delta > 0.3$), reaching a minimum of approximately $4 \times 10^{-11} M_{\odot} \text{ yr}^{-1}$ at $\delta = 0.40$, effectively approaching a `dead wind'' state.

\begin{figure}
   \centering
   \includegraphics[width=\hsize]{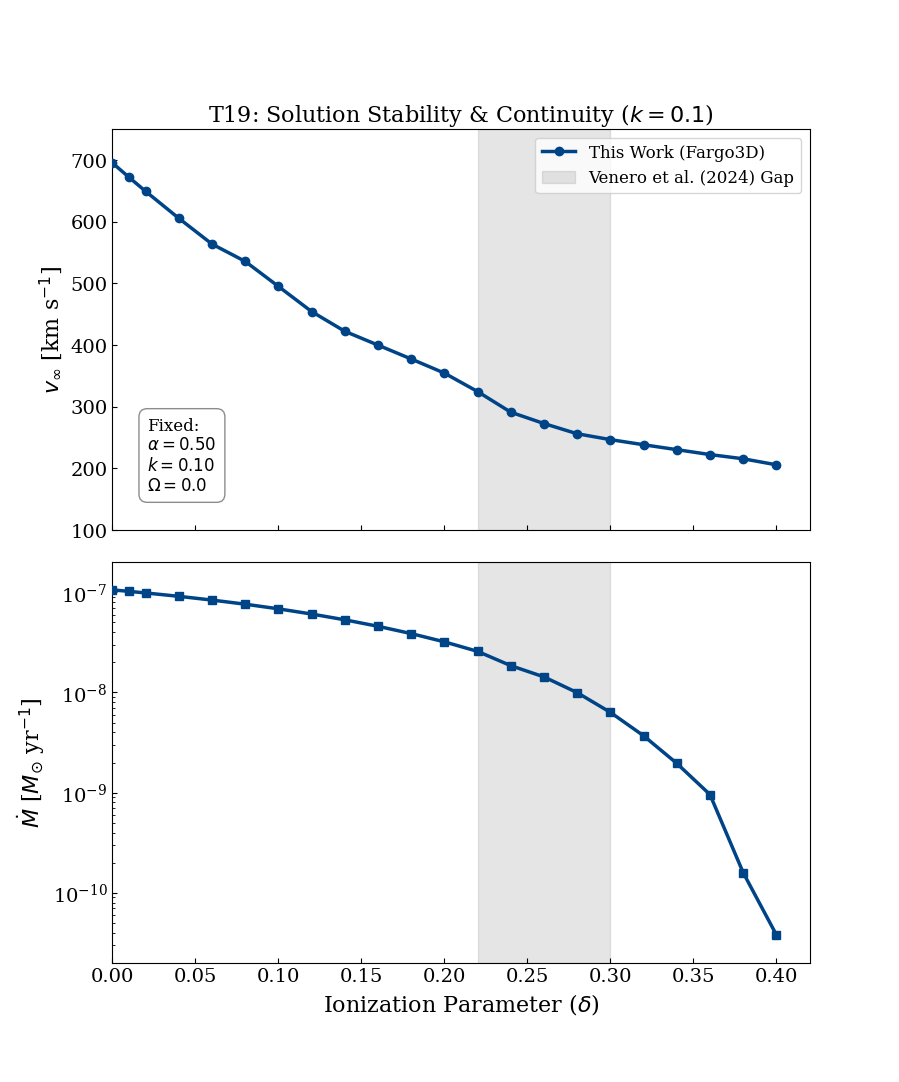} 
   \caption{Wind structure for the weak driving regime ($k=0.1, \Omega=0.0$). Top panel: Terminal velocity vs. $\delta$. Note the smooth inflection in the slope inside the grey region (the gap reported by \citealt{Venero+2024}). \textit{Bottom panel:} Mass-loss rate vs. $\delta$. The simulation captures the steep decline in $\dot{M}$ associated with the structural discontinuity in weak winds.}
   \label{fig:k01_structure}
\end{figure}

To understand the transition, Fig.~\ref{fig:k01_structure} details the wind structure for the non-rotating case ($\Omega=0.0$) across the region identified as a gap in stationary models ($0.22 \le \delta \le 0.30$). Our simulations reveal that the velocity profile (top panel) develops a distinct change in slope within the transition zone. While the mass-loss rate drops sharply, the transition is continuous, bridging the fast solution and the tenuous slow solution.

Regarding the convergence of the weak driving models ($k=0.1$), we investigated how the choice of initial conditions affects the final steady state. As shown in the sensitivity tests presented in Appendix \ref{sec:appendix_snapshots}, models initialized with different velocity gradients ($\beta=1.0$ and $\beta=2.0$) reach stable but numerically distinct stationary branches. Although the discrepancy in the final terminal velocity is small ($\sim 4\%$), its persistence over several years of evolution indicates that the time-dependent relaxation in this regime can isolate different stable equilibria within the same physical parameter space. This suggests that the final steady-state configuration near the transition regions of the m-CAK theory is slightly influenced by the initial kinematic state ($\beta$ parameter), a feature that we have consistently accounted for by using adaptive initialization as described in Section \ref{sec:methods}.

\section{Discussion}
\label{sec:discussion}

\subsection{The physical nature of the ``Gap'': Topology vs. Instability}
\label{sec:disc_gap}

A central result of this work is the demonstration that the ``forbidden regions'' or gaps in the $(\Omega, \delta)$ parameter space, previously reported in stationary hydrodynamic studies \citep[e.g.,][]{ Venero2016}, do not represent an absence of solutions. Instead, our time-dependent simulations reveal that these domains are populated by stationary winds characterized by a complex, non-monotonic acceleration profile featuring a distinct velocity plateau.

As shown in \S~\ref{sec:topology} (Fig.~\ref{fig:comparison_t19}), the transition from the fast to the slow wind regime is mediated by the development of a velocity plateau or ``kink'' in the supersonic region ($r \approx 2-10\,R_\star$). This feature represents a hydrodynamic adjustment where the flow locally maintains a nearly constant speed in response to the rapid decline of the radiative force term, which is proportional to $\rho^\delta$.

We suggest that the inability of the stationary code \texttt{HYDWIND} to find solutions in this domain is closely linked to the numerical difficulty of resolving the complex flow topology. While our time-dependent approach finds a stable bridge, the emergence of this complex acceleration profile hints at a potentially unstable physical landscape. In general, stationary methods rely on integrating the momentum equation from the sonic point outwards, typically searching for a specific regularity condition at a single critical point. The appearance of an extended plateau implies that the velocity gradient $\partial v/\partial r$ approaches zero in the supersonic flow. This topology likely creates multiple critical points or mimics a singularity, causing standard integration algorithms to diverge or fail to select a unique path.

In contrast, the time-dependent approach allows the system to relax naturally into this configuration. The flow navigates the complex force balance by adjusting its density structure (as seen in Fig.~\ref{fig:topology_structure}) to maintain mass flux conservation. However, the emergence of such a complex flow structure (a deceleration or stagnation followed by re-acceleration) indicates the wind occupies a \textit{metastable} equilibrium. This implies the solution is maintained in a fragile state between two competing stable regimes, characteristic of a domain with multiple solutions.

While our simulations converge to a single steady-state solution under the current initial conditions, the inflection in the velocity curve hints at a richer hydrodynamic landscape than previously assumed. This finding warrants further investigation to fully characterize the potential coexistence of multiple solution branches in this transition region, similar to the bi-stability reported in the rotational domain by \citet{Araya+2018}.

\subsection{Boundary conditions and the mass-loss rate discrepancy}
\label{sec:disc_boundary}

While our results show exact identity agreement with the stationary predictions for the terminal velocity in the fast wind regime, a significant qualitative divergence arises in the mass-loss rate behavior within the slow wind regime ($\delta \gtrsim 0.30$). As illustrated in the bottom panel of Fig.~\ref{fig:comparison_t19}, the stationary models of \citet{Venero2016} predict a sudden, exponential increase or ``jump'' in $\dot{M}$ immediately after the gap. This theoretically predicted bi-stability jump is expected to occur as the effective temperature decreases, triggering a shift in the ionization balance of the wind. In contrast, our time-dependent simulations yield a moderate, monotonic increase in $\dot{M}$.

We attribute this discrepancy to the different boundary conditions applied at the stellar surface. Standard stationary methods typically fix the photospheric optical depth to a constant characteristic value (e.g., $\tau_\star = 2/3$) to determine the density at the stellar surface. In the slow wind regime, where the velocity gradient is shallow, the density scale height increases. To maintain a constant integrated optical depth $\tau_\star \propto \int \rho \, dr$ under these conditions, the stationary solver is forced to artificially increase the base density, leading to the reported jump in the mass-loss rate.

In our time-dependent implementation, we adopt a fixed density boundary condition at the stellar surface ($\rho(R_\star) = \rho_0$), which effectively treats the photosphere as a stable hydrostatic boundary determined by the stellar structure, independent of the wind dynamics. Under this condition, the mass-loss rate is determined self-consistently. As $\delta$ increases, the changing force balance allows for a slightly higher mass flux, but the fixed density boundary acts as a regulator, leading to a stable mass loss evolution instead of the abrupt runaway predicted by stationary codes under different boundary assumptions.

In this context, it is worth noting that recent grid-based spectroscopic analyses based on stationary wind models may also be affected by similar limitations. In particular, the ISOSCELES project \citep{Araya2025} derives mass-loss rates using stationary hydrodynamic solutions. While this approach represents a major advance in quantitative spectroscopy, the use of stationary wind structures implies that the inferred mass-loss rates may be systematically overestimated in regimes where the flow is intrinsically non-stationary or where the velocity gradient becomes shallow. Our results suggest that part of the discrepancy between stationary and time-dependent mass-loss predictions may therefore be methodological in origin, rather than reflecting genuine differences in the underlying radiative driving.

Our finding of a continuous and monotonic behavior in $\dot{M}$ across the transition region is in excellent agreement with the latest empirical constraints across various metallicity environments. \citet{Verhamme+2024}, analyzing a sample of LMC supergiants, demonstrated that the theoretically predicted bi-stability jump is absent in observational data. This is further supported by recent quantitative spectroscopic surveys of Galactic B-supergiants \citep{deBurgos2024} and their counterparts in the Small Magellanic Cloud \citep{Bernini-Peron2024}, both of which report a smooth transition in wind properties rather than the abrupt mass-loss increase predicted by stationary models. These multi-environment observations consistently support the structural continuity naturally recovered by our time-dependent simulations.

\subsection{Implications for high-temperature stars}
\label{sec:disc_highT}

For the hottest model investigated (T25, $T_{\rm eff}=25$ kK), previous stationary works reported exclusively fast wind solutions, failing to converge to a slow branch. Our global maps (Fig.~\ref{fig:heatmaps_global}) confirm that stable solutions exist throughout the parameter space, including at high $\delta$.

However, our results clarify why a distinct ``slow'' regime was elusive in previous studies. Even at the highest ionization parameter considered ($\delta=0.40$), the wind velocity for T25 remains relatively high ($v_\infty \gtrsim 1000$ km s$^{-1}$), which is still significantly higher than the values obtained for cooler supergiants at similar $\delta$. Unlike the cooler models (T15, T19), where the radiative force weakens enough to allow winds with terminal velocities as low as $172$ and $254$ km s$^{-1}$ respectively, the intense luminosity of early B-supergiants maintains a strong driving force even when penalised by high $\delta$. This suggests that the hydrodynamic transition between fast and slow winds is temperature-dependent, and for stars hotter than $T_{\rm eff} \sim 23$~kK, the ``slow'' branch is effectively a moderately fast wind, making the structural distinction less pronounced than in cooler supergiants.

\subsection{Stability in the weak driving limit}
\label{sec:disc_weak}

Our exploration of the weak driving regime ($k=0.1$, see \S ~\ref{sec:weak_driving}) confirms the physical trend reported by \citet{Venero+2024}, where the mass-loss rate exhibits an inverse dependence on ionization. This contrasts sharply with the standard regime ($k \approx 0.32$), where higher $\delta$ generally sustains or increases the mass flux.

A key finding in this regime is the sensitivity of the solution to the initial conditions. We found that accessing the specific ``slow'' branch predicted by stationary theory required initializing the flow with a softer velocity gradient ($\beta=2.0$). This dependency suggests that the weak driving regime is characterized by multiple hydrodynamic solutions, where the history of the flow determines the final steady state. Furthermore, for $\delta \gtrsim 0.30$, the mass-loss rates drop to $\approx 4 \times 10^{-11} M_{\odot} \text{ yr}^{-1}$, implying that while these solutions are mathematically stable, they physically represent a ``dead wind'' limit. Our results show that the m-CAK solution space for weak driving is strictly continuous, connecting the fast branch with an ultra-tenuous slow branch that was previously unattainable with stationary methods.

\section{Conclusions}
\label{sec:conclusions}

We have presented a new time-dependent hydrodynamic study of radiation-driven winds in B-supergiants using the \textsc{Fargo3d} code. By exploring the full $(\Omega, \delta)$ parameter space for three representative stellar models (T15, T19, and T25), we have revisited the stability and existence of m-CAK wind solutions in regions previously identified as "forbidden" by stationary analyses. Our main conclusions are as follows:

\begin{enumerate}
    \item The ``Gap'' and solution existence: Our time-dependent simulations successfully find stationary wind solutions throughout the entire parameter space, including the previously reported gaps. This indicates that these ``forbidden regions'' are accessible via numerical methods capable of navigating complex flow topologies. However, the distinct kinematic features observed within these transition regions suggest they may represent a specific branch of a system characterized by multiple hydrodynamic solutions, without ruling out the existence of physical instabilities or alternative non-stationary states in these domains.
    
    \item Topological continuity: The transition between the fast and slow wind regimes is smooth but structurally complex. In the transition region (the former gap), the wind develops a distinct topology characterized by an extended velocity plateau or ``kink'' in the supersonic flow ($r \approx 2-10\,R_\star$). This kinematic structure is accompanied by a density profile that remains consistent with the velocity regime to satisfy mass conservation, allowing the flow to navigate the changing force balance without breaking stability—a feature that likely mimics a singularity in stationary shooting methods.
    
    \item Resolution of the mass-loss discrepancy: By adopting a fixed density boundary condition at the photosphere, which we argue is physically more robust than fixing the optical depth, we eliminate the artificial discontinuity in the mass-loss rate reported in stationary m-CAK models. It is important to distinguish this behavior, which arises from the mathematical sensitivity of the $\dot{M}$ vs. $\delta$ relation in the slow-wind regime \cite{Venero2016}, from the classical bi-stability jump in the mass-loss rate typically predicted as a function of decreasing effective temperature \citep{Pauldrach1990, Vink1999}. Our models predict a smooth and bounded variation in $\dot{M}$ across the $\delta$ parameter space, avoiding the abrupt exponential growth found in previous stationary studies. We note that the specific trend depends on the stellar parameters; while hotter models show a moderate increase, the coolest model (T15) at low rotation rates exhibits a moderate decrease in mass loss as $\delta$ increases.

    \item High-temperature limit: For early B-supergiants (T25), we confirm that while solutions exist at high $\delta$, they remain kinematically fast ($v_\infty \gtrsim 1000$ km s$^{-1}$) due to the intense radiative driving, explaining the historical absence of a distinct slow branch in stationary searches for these stars.

    \item Weak driving and initial conditions: In the weak driving regime ($k=0.1$), we confirm the inverse relationship where $\dot{M}$ decreases with increasing $\delta$. Our simulations show that the final steady state in this regime depends on the initial velocity profile ($\beta$), supporting the existence of multiple hydrodynamic solutions. While continuous stable solutions exist across the transition, the slow branch ($\delta > 0.3$) effectively represents a near-vanishing of the wind ($\dot{M} \approx 4 \times 10^{-11} M_{\odot} \text{ yr}^{-1}$).
\end{enumerate}

These results suggest that the m-CAK theory admits stationary solutions across a wider parameter range than previously identified by stationary methods. While our time-dependent approach provides a stable "bridge" between regimes, the emergence of complex topologies in the transition region leaves open the possibility of multiple steady-state solutions. This discovery warrants further investigation into the potential co-existence of multiple hydrodynamic solutions, where perturbations could trigger switches between stable states or lead to the physical instabilities observed in massive star winds.

\appendix
\section{Time Evolution and Initial Condition Dependency}
\label{sec:appendix_snapshots}

\subsection{Dynamical Relaxation to Steady State}

\begin{figure}[htbp]
    \centering
    \includegraphics[width=0.65\textwidth]{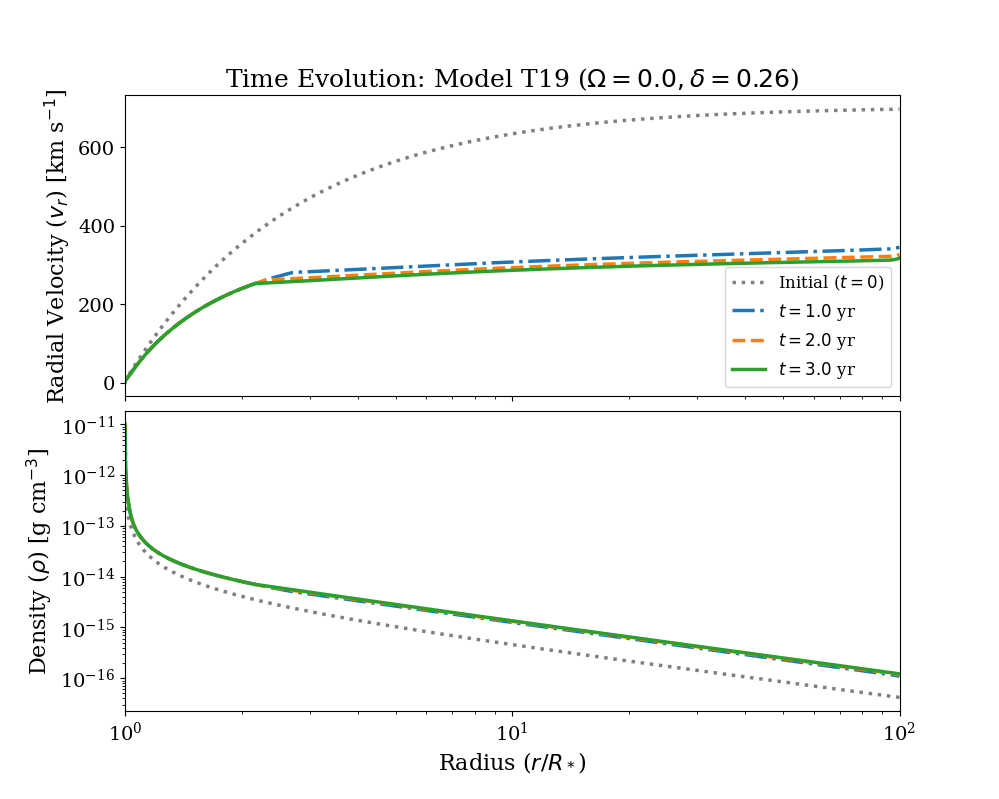}
    \caption{Time evolution of the wind structure for model T19 ($\Omega=0.0$, $\delta=0.26$). Top panel: Radial velocity profiles. Bottom panel: Mass density profiles. The dotted gray line represents the initial state ($t=0$). The intermediate snapshots (dashed lines) demonstrate the dynamic adjustment. The solid green line shows the final steady-state at $t=3.0$ years, where the characteristic velocity plateau is established.}
    \label{fig:A1_time_evolution}
\end{figure}

To demonstrate the dynamical relaxation of the wind from the initial conditions to the final steady state, we present snapshots of the radial velocity and density profiles at different dynamical times. Figure \ref{fig:A1_time_evolution} illustrates the temporal evolution of model T19 ($\Omega=0.0$, $\delta=0.26$). The flow starts from a standard $\beta=1.0$ velocity law and dynamically develops the characteristic velocity plateau ("kink") over several flow times. By $t=3.0$ years, the wind has fully evacuated the initial conditions across the entire $100 R_*$ domain, reaching a robust and stable stationary configuration.

\subsection{Sensitivity to Initial Conditions in the Weak Driving Regime}

\begin{figure}[htbp]
    \centering
    \includegraphics[width=0.75\textwidth]{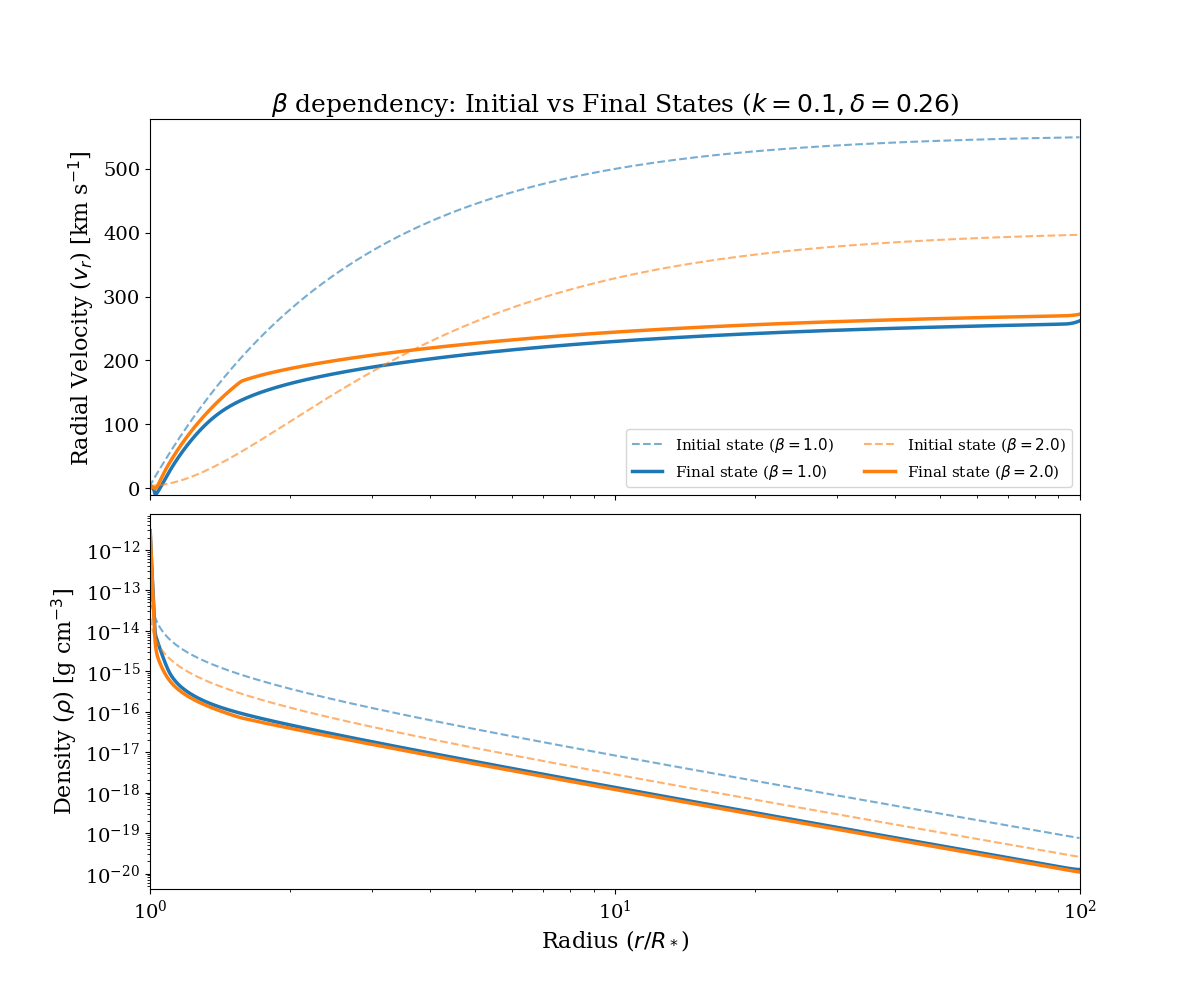}
    \caption{Initial condition dependency in the weak driving regime ($k=0.1, \Omega=0.0, \delta=0.26$). Top panel: Radial velocity profiles. Bottom panel: Mass density profiles. Dashed lines represent the initial states at $t=0$ for $\beta=1.0$ (blue) and $\beta=2.0$ (orange). Solid lines show the corresponding final steady states after $t=4.0$ yr. The model initialized with a faster profile relaxes to a slower stationary state, while the one starting from a slower profile reaches a faster state, confirming the existence of multiple stable solutions.}
    \label{fig:beta_dependency}
\end{figure}

To assess the sensitivity of the weak driving regime ($k=0.1$) to the initial kinematic state, we performed a comparison using two different acceleration profiles for model T19 at $\delta=0.26$. While the standard driving regime typically converges to a unique solution, the low line-driving limit exhibits a slight dependency on the initial flow history. Figure \ref{fig:beta_dependency} illustrates the evolution from the initial states to the final steady configurations.

Although both simulations converge to stable solutions, they settle into numerically distinct branches with a discrepancy that remains persistent after several years of evolution:
\begin{itemize}
    \item The model initialized with $\beta=1.0$ ($v_{\text{ini}} \approx 549$ km s$^{-1}$) relaxes to a steady state with $v_{\infty} = 260.67$ km s$^{-1}$ and $\dot{M} = 1.56 \times 10^{-8} M_{\odot}$ yr$^{-1}$.
    \item The model initialized with $\beta=2.0$ ($v_{\text{ini}} \approx 396$ km s$^{-1}$) converges to a branch with a terminal velocity of $271.75$ km s$^{-1}$ and a mass-loss rate of $1.43 \times 10^{-8} M_{\odot}$ yr$^{-1}$.
\end{itemize}

The difference in terminal velocity ($\sim 4\%$) and mass-loss rate ($\sim 9\%$) indicates that the system admits a discrete set of stable solutions within the same physical parameter space ($T_{\text{eff}}$, $\log g$, $k$, $\alpha$, $\delta$, and $\Omega$), where the selection of the final steady-state branch is determined by the initial kinematic condition (prescribed by the $\beta$-law) rather than by the underlying stellar parameters. This behavior suggests that the time-dependent attractor in the weak driving limit can isolate different topological branches depending on the initial energy of the flow, a feature that might be overlooked by stationary integration methods.

\begin{acknowledgments}
      We thank the anonymous referee for their constructive comments that helped to improve the quality of this paper. M.M. and M.C. acknowledge financial support from FONDECYT Regular 1241818. M.C. and I.A. acknowledge financial support from FONDECYT Regular 1230131 and 1261498. M.C. also acknowledges support from the Centro de Astrofísica de Valparaíso (CAV), CIDI N. 21 (Universidad de Valparaíso, Chile). This project has been partially co-funded by the European Union, Project 101183150 - OCEANS. Powered@NLHPC: This research was partially supported by the NLHPC's supercomputing infrastructure (CCSS210001).
\end{acknowledgments}

\bibliography{sample701}{}
\bibliographystyle{aasjournalv7}



\end{document}